\documentclass[journal]{IEEEtran}
\usepackage{amsmath,amsfonts,xcolor,cite,graphicx,booktabs,multirow,rotating}
\usepackage{algorithmicx,algorithm, array}
\usepackage[caption=false,font=normalsize,labelfont=sf,textfont=sf]{subfig}
\usepackage{textcomp, stfloats, url, verbatim}
\usepackage[dvipsnames]{xcolor}

\begin{document}

\title{Disaggregated Architectures and the Redesign of Data Center Ecosystems: Scheduling, Pooling, and Infrastructure Trade-offs}

\author{Chao Guo, Jiahe Xu, Moshe Zukerman, \IEEEmembership{Life Fellow, IEEE}
\thanks{\textit{This paper has been accepted for publication in IEEE Communications Magazine.
This is the author’s accepted manuscript version.} This work is supported in part by the Hong Kong Innovation and Technology Commission (InnoHK Project CIMDA) and in part by a grant from the Research Grants Council of the Hong Kong Special Administrative Region, China (CityU 11200523). (\textit{Corresponding author: J. Xu})}
\thanks{C. Guo is with the Centre for Intelligent Multidimensional Data Analysis Limited, Hong Kong SAR, China. (email: chaoguo6-c@my.cityu.edu.hk.)}
\thanks{J. Xu and M. Zukerman are with the Department of Electrical Engineering, City University of Hong Kong, Hong Kong SAR, China. (email: jiahexu3-c@my.cityu.edu.hk; moshezu@cityu.edu.hk).}
}

\maketitle

\begin{abstract}
Hardware disaggregation seeks to transform Data Center (DC) resources from traditional server fleets into unified resource pools. Despite existing challenges that may hinder its full realization, significant progress has been made in both industry and academia. In this article, we provide an overview of the motivations and recent advancements in hardware disaggregation. We further discuss the research challenges and opportunities associated with disaggregated architectures, focusing on aspects that have received limited attention. We argue that hardware disaggregation has the potential to reshape the entire DC ecosystem, impacting application design, resource scheduling, hardware configuration, cooling, and power system optimization. Additionally, we present a numerical study to illustrate several key aspects of these challenges.
\end{abstract}
\begin{IEEEkeywords}
Data center, hardware disaggregation, resource management, cooling and power management
\end{IEEEkeywords}

\section{Introduction} 
Data Centers (DCs), serving as the computing, storage, and exchange hub for global internet data, can be regarded as the brain and heart of the internet. 
They are often described as large IT resource pools, highlighting their extensive resource capacity, especially in cloud DCs. However, unlike a traditional pool of water where the amount can be continuously adjusted, the resources in a DC are composed of independent and isolated servers. Resource management and allocation are constrained by the physical boundaries of these server enclosures, limiting the flexibility of resource management. Over time, these limitations have led to increasingly significant challenges, including performance bottlenecks, resource underutilization, and reduced upgrade flexibility \cite{shan2018legoos, li2023pond}.

\begin{figure}
    \centering    \includegraphics[width=\linewidth]{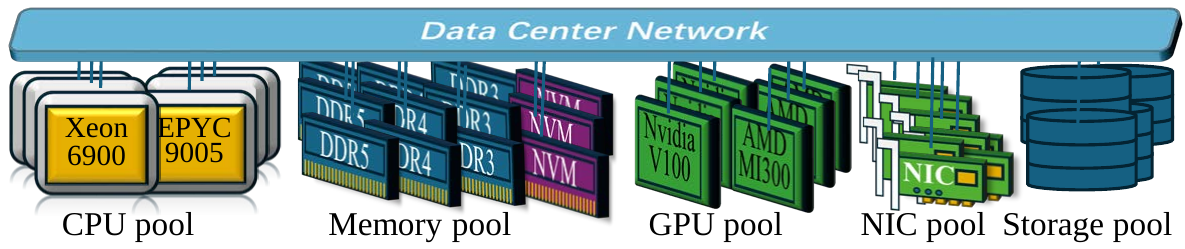}
    \caption{An idealized disaggregated architecture.}
    \label{fig:idealized_architec}
\end{figure}

Hardware resource disaggregation aims to overcome the limitations of traditional server architecture by restructuring IT resources in DCs into shared resource pools, as illustrated in Fig.~\ref{fig:idealized_architec}. This approach enables the dynamic selection of resource blocks to configure tailored systems, significantly improving resource flexibility and utilization. Additionally, the decoupling of different resource types allows for independent management, enabling agile and cost-efficient hardware upgrades. Moreover, disaggregation facilitates efficient management of heterogeneous resources, which is particularly beneficial as new resource types, such as diversified accelerators (e.g., GPUs, Data Processing Units (DPUs), and Neural Processing Units (NPUs)), continue to emerge. Service reliability can also be improved, as disaggregation provides finer failure granularity; for example, a CPU can remain operational even if a memory module fails \cite{lin2020disaggregated}.

The realization of these benefits requires both fast network support and software-level evolution. Much effort adheres to the transparency principle, which aims to hide underlying hardware changes from applications to maintain compatibility with existing software (e.g., Pond \cite{li2023pond}). Other research efforts have focused on leveraging the features of hardware disaggregation to improve application performance through software-level evolution. 
For instance, memory sharing can enhance the performance of distributed and parallel computing applications by allowing shared data to reside in a shared memory space, enabling multiple hosts to access it directly without data copying or transfer between these hosts \cite{angel2020disaggregation}. Network evolution has also made significant progress, with Compute Express Link (CXL)---an interconnect technology built on PCIe---emerging as the most popular solution for disaggregated architectures. Despite these advancements, the fully disaggregated architecture shown in Fig.~\ref{fig:idealized_architec} remains infeasible across an entire large-scale DC. As a result, several compromise approaches have been adopted: (1) limiting the scale of disaggregation to small groups of hosts \cite{li2023pond}, one or several racks \cite{katrinis2016rack}, or a cluster \cite{zervas2018optically}; and (2) reserving a small amount of local memory within each CPU node (or host) for fast access, mitigating the stringent communication demands between CPUs and memory.
Despite these compromises, disaggregation has already proven effective in addressing the aforementioned challenges, driving significant interest and investment from both industry and academia, including Microsoft Azure \cite{li2023pond}, Meta \cite{maruf2023tpp}, NVIDIA \cite{maruf2023tpp}, and Intel \cite{zhong2024managing}.

Hardware disaggregation brings significant benefits along with new challenges and research opportunities. From a long-term perspective, this architectural revolution is poised to transform the entire ecosystem, impacting hardware manufacturing, cooling and power management, rack placement, and application development. In this article, we provide a brief overview of the motivations and recent advancements in hardware disaggregation across both industry and academia. We then outline key challenges and research opportunities that have received limited attention in existing studies. Finally, we present numerical results that highlight the challenges of resource distribution in DCs, emphasizing the importance of co-designing hardware organization (dimensioning) and resource scheduling.

\section{Motivations}
Hardware disaggregation is driven by various motivations, and this section introduces several key motivations.

\subsubsection{Scaling Bottleneck}
The tight coupling of resources in traditional server architectures makes independent resource scaling difficult. Resource capacities within a server are fixed by the physical constraints of the motherboard, often requiring the replacement of the entire server to scale a single resource, leading to high costs and inefficiencies.
The most significant scaling bottleneck occurs with memory expansion. Despite the increasing core count per CPU, e.g., the Intel Xeon 6 processor (as of March 2025) offering up to 288 cores \cite{Intel_Xeon_6}, memory bandwidth (data read/write rate) and capacity per core have failed to keep pace with the growth of CPU core. This is primarily due to the tight coupling between memory and CPUs in modern architectures.
Since CPUs integrate memory controllers within their chip space, memory scaling is constrained by CPU design limitations. Specifically, (1) a memory controller is only compatible with a specific memory type, e.g., DDR4 controllers cannot support DDR5 or Non-Volatile Memory (NVM), and (2) increasing memory bandwidth requires additional space and pins, complicating CPU design. As a result, while CPU performance continues to scale, memory expansion faces architectural bottlenecks. Disaggregated architectures, facilitated by technologies such as CXL, address these challenges by supporting more flexible resource pooling and allocation, improving scalability while considering interconnect constraints.

\subsubsection{Heterogeneity Management Challenges}
Resource types in DCs have expanded rapidly, with GPUs, NPUs, and DPUs becoming increasingly important alongside traditional resources like CPUs and memory. Even within the same resource type, different products cater to specific applications (e.g., NVIDIA V100 GPUs for AI training and T4 GPUs for inference \cite{Jin2024distmind}). 
Although these specialized components enhance performance, they also increase resource management complexity. Disaggregation simplifies this by enabling resource pooling and dynamic composability, improving flexibility and efficiency. 

\subsubsection{Resource Stranding}
Resource stranding occurs when one resource type in a server is fully utilized while others remain underused, leading to wasted capacity. This issue arises from mismatches between workload demands and hardware configurations. For instance, deep learning servers require large memory capacities for training but significantly less for inference, resulting in memory stranding. As reported in \cite{li2023pond}, memory stranding in Azure reaches 25\%, highlighting significant inefficiencies.

\subsubsection{Others}
Additional motivations include cost-effective upgrades and finer failure granularity. Disaggregation allows independent upgrades of components like compute, memory, and storage, reducing costs. It also enhances resilience, as the failure of one resource type does not necessarily affect others.

\section{Progress}
Multiple solutions have been proposed for interconnecting disaggregated resources, with optical interconnects \cite{zervas2018optically}, Remote Direct Memory Access (RDMA) \cite{Jin2024distmind}, Gen-Z \cite{hong2020hardware}, and CXL \cite{li2023pond} being the most prominent. Optical interconnect technologies leverage optical circuit switching, optical packet switching, space division multiplexing, and wavelength division multiplexing to enable flexible, high-bandwidth, and low-latency communication. RDMA facilitates direct memory access between servers, bypassing the operating system and CPU, thereby improving access speed. This technology has matured significantly and is implemented in InfiniBand and RDMA over Converged Ethernet (RoCE), making it a popular choice in high-performance computing environments. Gen-Z is a memory-centric interconnect technology that provides scalable, low-latency access to a wide range of memory and other resources across disaggregated systems. It is compatible with PCIe. Unlike traditional PCIe-based interconnects, Gen-Z supports memory-semantic communication, enabling devices to directly access remote memory without CPU intervention.

In 2019, Intel introduced CXL, an interconnect technology built on top of PCIe. Unlike Gen-Z, CXL adds cache coherence support, which eliminates the need for data copying and synchronization between devices, thereby significantly reducing latency and improving access speed \cite{das2024introduction}. CXL supports the connection of a CPU node to a wide range of devices, including different types of memory modules, NVM, accelerators, smart NICs, and solid-state drives (SSDs) \cite{das2024introduction}. With this capability, a CPU node with a specific memory controller (e.g., DDR5) can use different memory types (e.g., DDR4 and NVM) for memory expansion. As a result, older DIMMs from decommissioned machines can also be reused, enhancing cost-efficiency and sustainability. These advantages make CXL the most popular interconnect for disaggregated architectures, which has attracted significant support from dozens of industry giants, including NVIDIA, AMD, Microsoft, Google, and Huawei. In 2022, CXL absorbed Gen-Z technology, further solidifying its position as the leading interconnect standard.

Recent progress in hardware disaggregation has primarily focused on CXL-based solutions, with particular emphasis on memory disaggregation due to its significant bottleneck. CXL achieves a latency of 170--250 ns, which is higher than (though close to) the 80--140 ns latency of DDR memory access \cite{aguilera2023memory}. This difference is primarily due to CXL's use of serial transmission and a request-response model, which requires protocol encapsulation and processing, incurring additional overhead. To compensate for this latency overhead, CXL-based memory disaggregation solutions typically retain a certain amount of DDR local memory on each CPU node. This approach mitigates the performance slowdown caused by the latency overhead of CXL. An example of this strategy is Pond, a CXL-based memory disaggregation platform proposed by Microsoft Azure in 2020 \cite{li2023pond}. In Pond, every 8 to 16 CPU nodes share a memory pool. Each VM is assigned a combination of local and remote memory. Machine learning is used to determine the appropriate amount of remote memory allocated to a VM, ensuring that application performance is not compromised. 
In addition to local memory reservation, memory tiering has been explored to further enhance access speeds. The core idea is to place the most frequently accessed data (hot pages) in local memory while keeping less frequently accessed data in remote memory. Memory tiering solutions can be categorized into software-based and hardware-based approaches. Software-based solutions rely on the hypervisor or operating system for management, with representative examples including TPP, proposed by Meta and NVIDIA. Hardware-based solutions, such as Intel’s Flat Memory Mode \cite{zhong2024managing}, delegate management to the CPU processor or memory controller. 

Apart from memory, PCIe/CXL is the primary interconnect for disaggregating and pooling resources such as accelerators, NICs, and SSDs, as demonstrated in commercial solutions like GigaIO Fabrex (https://gigaio.com/products/fabrex-system-overview/) and the Liqid platform (https://www.liqid.com). This choice is primarily driven by PCIe/CXL’s high bandwidth and low latency, which are well-suited for emerging AI and data-intensive workloads. Moreover, these resources are already connected via PCIe in conventional server architectures, ensuring full compatibility. Modern servers already use PCIe Gen 4.0 and Gen 5.0 to connect AI accelerators like GPUs, delivering 32 GB/s (1 GB = 8 Gbps) and 64 GB/s (x16, bidirectional) bandwidth with tens to hundreds of nanoseconds latency. Accordingly, using CXL 3.0 and PCIe 7.0, offering up to 128 GB/s and 512 GB/s with even lower latency, can easily accommodate future communication demands in disaggregated architectures. In addition, Optical I/O is being integrated with PCIe/CXL to scale disaggregated systems. For instance, Ayar Labs offers an optical I/O solution supporting both PCIe and CXL, delivering up to $\sim$8 Tbps ($\sim$1 TB/s) bidirectional bandwidth with $\sim$10 ns latency (excluding fiber delay), enabling effective interconnects across racks without repeaters. Recently, RDMA technologies like InfiniBand NDR now reach 50 GB/s (400 Gbps), which is comparable to PCIe 5.0. However, with latencies in the microsecond range, RDMA is less suitable for latency-sensitive components. Still, it can be a cost-effective option for connecting lower-performance devices such as SSDs. 

\section{Challenges and Research Opportunities}
\begin{figure*}
    \centering
    \includegraphics[width=0.8\linewidth]{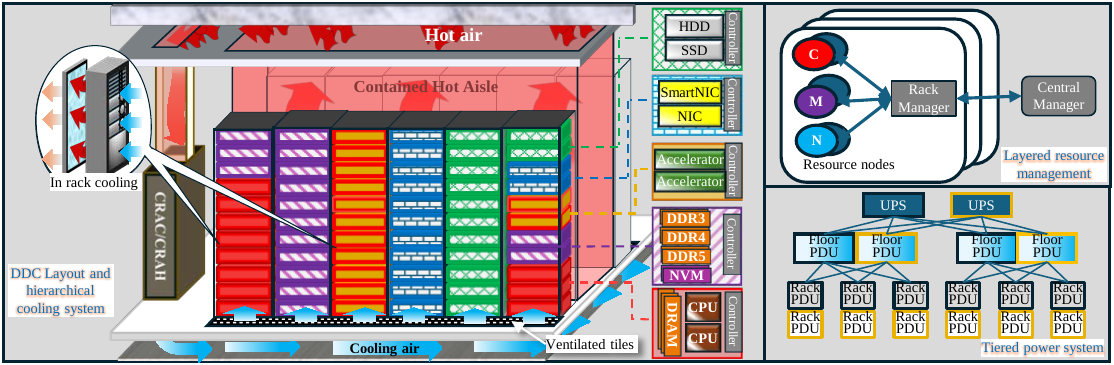}
    \caption{Cooling, power, and resource management for a DDC.}
    \label{fig:cooling_power_manager}
\end{figure*}

This section discusses the challenges and research opportunities that have received limited attention in existing studies.

\subsection{Pool Configuration Design}
\label{sec:chal_config}
Most existing disaggregation architectures focus on scales smaller than an entire DC, with rack-scale disaggregation being the most widely studied. Rack-scale disaggregation treats one or several racks as a resource pool, where each task can utilize different resource types from a single pool, while different pools remain isolated. At a higher level, rack-scale disaggregation extends the traditional server boundary to encompass an entire rack (or multiple racks). However, this approach may inherit configuration issues from traditional servers, necessitating carefully designed pool configurations to maximize efficiency.

One configuration strategy is \textbf{function-specific pool design}, where resources are organized into pools tailored for specific workloads. This approach ensures that critical tasks, such as compute-intensive workloads with high CPU but low memory requirements, are adequately supported. However, it relies heavily on accurate demand forecasting to align pool configurations with actual workload patterns. Inaccurate predictions can lead to resource underutilization and inefficiency. An alternative approach is \textbf{uniform resource distribution}, where resources are evenly allocated across pools. 
While easy to implement and less dependent on demand forecasting, this strategy may struggle to handle demanding workloads that require a significant amount of a specific resource. 

Existing research on resource management primarily focuses on workload scheduling based on fixed hardware configurations. However, disaggregated architectures present a unique opportunity to \textbf{co-design resource scheduling and pool configuration} for optimal efficiency. By dynamically adjusting pool configurations based on real-time workload demands and scheduling decisions, the system can better adapt to varying resource requirements. For example, during periods of high compute-intensive workloads, the system could temporarily reconfigure pools to allocate more CPU resources, while prioritizing memory allocation during memory-intensive phases. This co-design approach not only improves resource utilization but also enhances the system's ability to handle diverse and dynamic workloads. Implementing such a strategy requires close integration between the resource management layer and the hardware configuration layer, enabling real-time feedback and optimization. This holistic approach represents a promising direction for future research in disaggregated architectures.

\subsection{Diversified Disaggregation Scale}
Different resource types can tolerate varying levels of disaggregation scale due to their distinct communication requirements. A unified disaggregation scale often leads to inefficiency, as it may not fully leverage the potential of each resource type. Rack-scale disaggregation is primarily designed for memory disaggregation. If all resources are constrained to the scale designed for memory disaggregation, the potential benefits of disaggregation will not be fully realized. 

A diversified disaggregation scale can be realized through a hierarchical pool design, where a large resource pool contains multiple nested small pools. The smallest pooling unit consists of CPU and memory nodes. Given the stringent latency and bandwidth requirements of memory access, each CPU is connected to multiple memory nodes within its pool via dedicated, direct CXL links to avoid performance degradation due to traffic contention or switching latency. Additionally, CXL-supported memory interleaving allows each CPU to access multiple memory nodes concurrently, thereby maximizing memory bandwidth. Multiple such small pools can be grouped to share a common set of accelerator and NIC nodes, forming a larger pooling unit. Since accelerators and NICs typically tolerate microsecond-level latency and lower bandwidth sensitivity, PCIe or CXL switches can be used in the interconnection to provide high connectivity. One or more of these larger pools may, in turn, share access to a set of storage devices (e.g., SSDs or HDDs), following the same principle.
For resource scheduling, a centralized controller maintains a global view of the hierarchical pool structure. Each CPU node is associated with a set of candidate memory, accelerator/NIC, and storage nodes. Upon receiving a composable system request, the controller first selects a CPU and then chooses additional components from its associated lists. To reduce the overhead of path discovery and mitigate performance fluctuations caused by network load, communication among the selected nodes follows predefined static paths, ensuring more stable system performance.

\subsection{Paradox: Simplified Pools vs. Increased Nodes}
When reviewing the literature, we encounter a paradox. Some researchers argue that resource disaggregation simplifies resource management and improves resource efficiency by creating a unified and shared resource pool, as opposed to managing independent devices \cite{shan2018legoos}. This approach enables dynamic resource allocation, theoretically eliminating or significantly alleviating the bin-packing problem. Conversely, many researchers contend that disaggregation complicates resource management. By separating different components from individual servers and exposing them to a shared network, the number of manageable nodes increases significantly \cite{zervas2018optically}. Additionally, network performance metrics such as latency and capacity, which were previously managed entirely by the motherboard, now become critical factors in resource management. These considerations make the scheduling problem within a disaggregated architecture more complex.

These seemingly contradictory views stem from different perspectives. Resource disaggregation indeed increases the number of nodes, as each resource type (e.g., memory, CPU, GPU) can be independently designed as scalable modules.
Resource allocation for traditional architectures involves selecting one server from \( n \) servers, with a complexity of \( O(n) \). In disaggregated architectures, a node must be selected for each of the \( R \) resource types, leading to a complexity of \( O(n^R) \).
From a high-level perspective, these resources can be abstracted into unified resource pools, creating the illusion that resource management becomes easier. However, achieving this illusion requires sophisticated resource allocation design to hide the complexity of the underlying hardware. This approach necessitates hierarchical resource management, as illustrated in the top-right part of Fig.~\ref{fig:cooling_power_manager}. In this design, resource allocation is partitioned into multiple layers, such as the node layer, rack layer, cluster layer, and central layer. This concept aligns with the network protocol stack principle, where each layer focuses only on its own responsibilities, and the lower layers provide necessary information (e.g., aggregated capacity) to the upper layers. In this way, the higher-level manager maintains a view of resource pools, while each pool manages the detailed information of its constituent nodes.

\subsection{How to Treat Traditional Servers?}
\label{sec:chal_serv}
Despite the numerous benefits of resource disaggregation, its performance generally falls short of that of integrated server architectures for most applications. Consequently, traditional servers are expected to dominate DCs for the foreseeable future, while disaggregated resources will gradually be integrated into these systems. The challenge lies in effectively managing the coexistence of disaggregated resources and traditional servers to maximize profitability. 

One approach is to treat them separately by establishing distinct clusters or zones for each architecture. This allows disaggregated resources to be reserved for specific use cases where they excel, such as big data analytics, AI training, or scenarios requiring maximized resource utilization for cost or efficiency reasons. Additionally, specialized management of security policies and performance optimizations can be tailored to leverage the strengths of each architecture.

An alternative approach is to integrate disaggregated resources with traditional servers, using them as a complement to address the limitations of current server architectures, such as resource expansion and resource stranding. For example, a server rack or cluster can be equipped with a certain amount of disaggregated memory resources to enable memory expansion and sharing. This approach, however, requires the development of advanced orchestration platforms capable of managing both disaggregated and traditional resources in a unified manner. Such platforms must handle the complexities of resource allocation, performance optimization, and fault tolerance across heterogeneous architectures.
\subsection{Power and Cooling Constrained Resource Placement}
Resource disaggregation introduces significant challenges in resource placement, particularly when considering power and cooling systems. 
In DCs, racks are typically arranged in rows to optimize cooling efficiency, space utilization, and operational convenience. A common approach is the hot/cold aisle containment design, as illustrated in Fig.~\ref{fig:cooling_power_manager}. The figure shows two rows of racks placed back-to-back, with hot aisles isolated using partitions to minimize mixing hot and cold air. The hot air is directed through overhead plenums and ducts to Computer Room Air Conditioning/Handling (RAC/CRAH) units, where it is cooled and circulated back to the racks via raised floors and ventilated tiles. This approach is also referred to as in-room cooling, providing relatively uniform cooling capacity for the racks across the room. Uniform cooling capacity requires balanced power consumption across these racks, creating a constraint for dynamic resource pooling. The concentration of homogeneous resources within specific racks leads to skewed power consumption and heat generation in each rack, which will lead to significant hot spots and low cooling efficiency under the in-room cooling conditions. 
To address this challenge, one approach is to adopt hybrid cooling solutions that combine in-room cooling with finer-grained cooling techniques, such as in-rack cooling (as shown in Fig.~\ref{fig:cooling_power_manager}) or direct-to-chip liquid cooling. 
While finer cooling granularity improves efficiency, it also increases cost and operational complexity, necessitating comprehensive design optimization. 

Power constraints further complicate resource placement. Modern DCs typically employ a three-tier power hierarchy, as depicted in the lower-right part of Fig.~\ref{fig:cooling_power_manager} \cite{baxi2025online}. At the top level, Uninterruptible Power Supply (UPS) units deliver power from the grid or generators to individual rooms. Floor Power Distribution Units (PDUs) distribute power within each room, and rack PDUs supply power to individual racks. These systems are designed with redundancy, including normal and failover capacity, to ensure reliability during power outages. In addition to optimizing resource placement under power system constraints, another research direction is restructuring the power system to accommodate skewed rack power consumption. Both directions present significant challenges and optimization opportunities.

\subsection{Other Aspects}
In a fully pooled system, a task could in theory draw resources from multiple nodes, but this is impractical for some types, like CPUs. While multi-CPU systems exist, spreading a VM across CPU sockets is rare due to complexity and overhead \cite{li2023pond}. Some memory disaggregation systems allow using multiple memory nodes, but this demands careful scheduling \cite{wang2024boosting}.
Another key concern is the local-to-remote memory ratio. For example, Intel uses a 50\% local memory threshold and dynamically adjusts allocations to prevent performance loss \cite{zhong2024managing}. However, such adjustments add allocation complexity and may fail when local memory is exhausted, requiring costly workload migration.

Virtualization technologies like SDN, containers, and VMs are widely used in data centers, and integrating them with disaggregated architectures is key to enabling software-defined infrastructure. One research direction is defining what disaggregated hardware information should be exposed to orchestrators like Kubernetes. Implementing transparent policies that conceal disaggregated features can support existing applications. Conversely, revealing certain features can empower application developers to enhance performance, for instance, leveraging shared memory to reduce data copying overhead among various subtasks of a job. Another underexplored area is Virtual Network Embedding (VNE), a core SDN challenge well-studied in traditional architectures but lacking research in disaggregated settings.

\section{Case Study}
In this section, we demonstrate how pool configuration impacts efficiency in a resource allocation scenario involving predefined requests and resource nodes (servers and disaggregated nodes). Nodes are categorized into pools based on different configuration principles.
To ensure fairness, we use Integer Linear Programming (ILP) assisted by a commercial solver (Gurobi) for optimal allocation. The objective is to minimize total resource usage, defined as the sum of resource capacities across active nodes. To eliminate the influence of units, we use normalized resource capacity values. Given the presence of multiple resource types, we adopt a weighted sum approach, with weights indicating the relative importance of each resource type. Each request is assigned one host node (a server or a disaggregated CPU node), multiple memory nodes, and at most one node of other disaggregated resource types. All nodes used by a request must be from the same pool. Additionally, the host node assigned to a request must provide local memory to meet the local memory threshold.

\begin{table}
\centering
\caption{Resource Demand Settings}
\begin{tabular}{p{2cm}|p{5.5cm}}
\hline
\textbf{Resource Type} & \textbf{Demand} \\ \hline
CPU cores  & ($N_{core}=$) \( \mathcal{N}(2, 6^2;[1,32]) \) \\ 
Memory (GB) & \(N_{core}\times \mathcal{N}(2, 4^2;[1, 12]) \) \\ 
Accelerator& 75\% requests with no demand while others with $N_{core} \times \mathcal{N}(0, 3^2; [1, 8]) $ units (capped at 32 units), among which two-thirds request GPUs.\\
Local memory threshold & Uniform(0, 0.5) $\times$ Total memory demand\\
\hline
\end{tabular}
\label{tab:demand}
\end{table}

\begin{table}[t]
    \centering
    \caption{Resource Node Settings}
    \begin{tabular}{l|l|c}
        \hline
        Node Type & Configuration & Qty.\\ 
        \hline
        CPU&32 cores + 64 GB memory &16  \\
        Memory& Total memory capacity of 1280 GB. &- \\
        GPU&32 GPU units &8  \\
        FPGA&32 FPGA units &4   \\
        S1 &32 (CPU) cores + 128 GB memory &4  \\
        S2 &32 cores + 64 GB memory &3  \\
        S3 &32 cores + 256 GB memory &2  \\
        S4 &32 cores + 128 GB memory + 32 GPU units &2  \\
        S5 &32 cores + 128 GB memory + 32 FGPA units&1  \\
        \hline
    \end{tabular}
    \label{tab:resource}
\end{table}

\subsection{Test Settings}  
Table~\ref{tab:demand} summarizes the request settings. We use the truncated normal distribution $\mathcal{N}(\mu, \sigma^2; [a, b])$, with mean $\mu$, variance $\sigma^2$, and bounds $[a, b]$, to model both individual resource demands and inter-resource ratios, capturing the long-tail behavior typical of data centers and reflecting diverse workloads from domains such as scientific computing, finance, and e-commerce. As shown, 25\% of requests are \textit{accelerator-assisted}. The remaining requests are categorized based on per-core memory demand: [1, 3] GB for \textit{compute-intensive}, [3, 6] GB for \textit{general-purpose}, and [6, 12] GB for \textit{memory-intensive} workloads.

Table~\ref{tab:resource} shows the node configuration settings, including four disaggregated resource types (CPU - FPGA) and five server node types (S1 - S5). Note that only the total memory capacity is listed, as individual node sizes are determined by pool configuration policies. 
We consider two ways to treat these servers and disaggregated resources: separately and mixed. In the separate approach, servers are outside the disaggregated resource pools. These disaggregated resources are grouped into four pools with three configuration policies considered as follows.

\textbf{C1 (Uniform):} Each pool has 4 CPU nodes, 2 memory nodes (160 GB each), 2 GPU nodes, and 1 FPGA node.  

\textbf{C2 (Function-Specific):} The four pools are configured as \textit{general}, \textit{compute-optimized}, \textit{memory-optimized}, and \textit{accelerator-assisted}, with per-core memory capacities of 4~GB, 2~GB, 8~GB, and 4~GB, respectively. CPU nodes are evenly distributed across the four pools. The general pool includes two 128-GB memory nodes, the memory-optimized pool two 384-GB memory nodes, and the accelerator-assisted pool two 128-GB memory nodes. All GPU and FPGA nodes are allocated to the accelerator-assisted pool.  

\textbf{C3 (Refined Function-Specific):} Similar to C2, this configuration has been optimized according to request distributions. For instance, in alignment with the fact that 25\% of requests require an accelerator and have an expected per-core memory demand of 4.5~GB, the accelerator-assisted pool is set up with 4 CPU nodes, 2 memory nodes (160~GB each), and all GPU and FPGA nodes. 
The general, compute-optimized, and memory-optimized pools are configured with: 5 CPU nodes and 2 memory nodes (197-GB each); 4 CPU nodes; 3 CPU nodes and 2 memory nodes (283-GB each).

In the mixed approach, these 12 servers are distributed into the four pools to use disaggregated resources. With the C1 policy, these 12 nodes are placed into the four pools as uniformly as possible. With the C2 and C3 policies, these 12 nodes are placed in the four pools that match their functions, e.g., the two memory-optimized servers are placed into the memory-optimized pool. Overall, there are six configurations: C1, C2, and C3 with separate or mixed servers. 

For function-specific configurations, requests are assigned to matching pools as close as possible, with penalties for mismatches. Penalties are set as 0 for perfect matches, 1 for a general request being placed in the compute/memory-optimized pool or a compute/memory-intensive request being placed in the general-purpose pool, and 2 otherwise. The objective function prioritizes minimizing penalties for these configurations, followed by minimizing total resource usage. 
The weights in the objective function to indicate the importance of different resource types are set at 100 for CPU, 10 for accelerators, and 1 for memory. 
We conducted 11 runs for each test condition and report the average results along with 95\% confidence intervals based on Student's t-distribution. The same set of random seeds was used across all test conditions to ensure consistency.

\subsection{Results}
\begin{figure}
    \centering
    \includegraphics[width=0.7\linewidth]{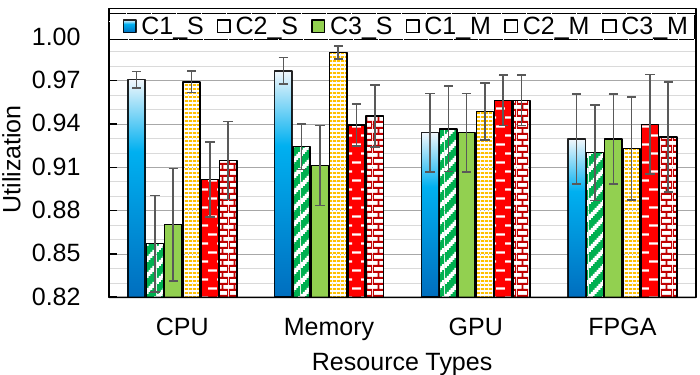}
    \caption{Resource utilization under different pool configurations.}
    \label{fig:utilization}
\end{figure}

\begin{figure}
    \centering
    \includegraphics[width=0.7\linewidth]{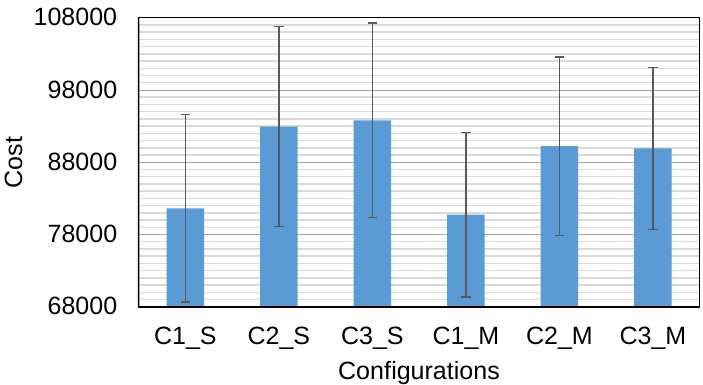}
    \caption{Cost under different pool configurations.}
    \label{fig:cost}
\end{figure}
We first evaluate the resource efficiency of different configurations. To evaluate system capacity, in the test, we start by feeding the ILP solver with a large number of requests and iteratively reducing them until all requests can be accommodated, ensuring resource saturation for assessing final utilization. 
Fig.~\ref{fig:utilization} shows the utilization of the four resource types under the six configurations, where C$k$\_S and C$k$\_M ($k=1,2,3$) correspond to C$k$ configuration principles with servers treated separately and mixed, respectively. 

Three observations are made from the CPU utilization results: 1) Uniform configuration (C1) achieves higher utilization than function-specific configurations (C2 and C3); 2) mixing disaggregated resources with traditional servers improves resource utilization compared to treating them separately; 3) the performance of function-specific configuration is sensitive to workload prediction accuracy, and higher prediction accuracy leads to better performance (C3 performs better than C2). These results are generally consistent with our analyses in Section~\ref{sec:chal_config} and Section~\ref{sec:chal_serv}. 

Memory utilization generally follows the same trends as CPU utilization, with one notable exception: C3\_S yields lower utilization than C2\_S.  This seems to suggest that function-specific configurations are not always highly sensitive to workload prediction accuracy. However, it is important to note that the prediction accuracy is statistical, so optimal performance is only expected when the problem scale (e.g., number of resources and requests) approaches infinity, which is not the case in our test setup.
The trends for GPU and FPGA results are less clear than for CPUs. For GPUs, we can still observe a clear advantage of the mixed strategy over the separate one. However, this distinction is not evident for FPGAs. This is partly due to the limited test scale, with only 8 GPU nodes and 4 FPGA nodes. Another key reason lies in the lower weight assigned to accelerators compared to CPUs in the optimization objective. These results reflect the trade-offs in balancing multiple resource types.

We next evaluate the cost-efficiency of different configurations by processing the same request volume (50 requests) and calculating the total cost of the physical resources used (i.e., all active nodes). Based on market analysis, we assign normalized unit costs (relative to memory) as follows: 100 per CPU core, 1 per GB of memory, 300 per GPU unit, and 100 per FPGA unit.
The results show that the uniform configuration achieves the highest cost-efficiency, and mixing disaggregated resources with traditional servers further improves efficiency. The sensitivity of function-specific configurations to workload prediction accuracy is evident when comparing C3\_M with C2\_M, but not when comparing C3\_S with C2\_S. This exception has already been explained in the utilization analysis and is not repeated here.

However, this case study has some limitations, including a small test scale, due to our reliance on ILP for fair comparison, and the omission of performance metrics like system response time. Nonetheless, the results highlight the importance of pool configuration in achieving high performance.

\section{Conclusion}
Hardware disaggregation reshapes DCs, enhancing flexibility and efficiency but introducing systemic challenges. Our analysis identifies key trade-offs in resource scheduling, dynamic pooling, and infrastructure co-design. While advances in CXL and tiered memory help, gaps remain in managing interdependencies among resources, cooling, and energy. Future research should focus on cross-layer optimization to integrate disaggregation with physical constraints, enabling adaptive resource orchestration for balanced performance and sustainability.

\bibliographystyle{IEEEtran}
\bibliography{refs.bib}

\begin{IEEEbiographynophoto}{Chao Guo} is currently a Postdoctoral Fellow with the Centre
for Intelligent Multidimensional Data Analysis Ltd., Hong Kong Science Park, Hong Kong. His research interests include the areas of data center resource scheduling and network optimization, cable path planning, and wireless networking.
\end{IEEEbiographynophoto}

\begin{IEEEbiographynophoto}{Jiahe Xu} received a B.Eng. degree at the Harbin Institute of Technology and an M.Sc. degree at the City University of Hong Kong, where she is currently pursuing a Ph.D. degree with the Department of Electrical Engineering. Her research interests include network optimization, data center resource management, and network slicing. 
\end{IEEEbiographynophoto}

\begin{IEEEbiographynophoto}{Moshe Zukerman}(M’87–SM’91–F’07–LF’20) received the B.Sc. degree in industrial engineering and management, the M.Sc. degree in operations research from the Technion – Israel Institute of Technology, Haifa, Israel, and the Ph.D. degree in engineering from University of California, Los Angeles, in 1985. He was an independent consultant with the IRI Corporation and a Postdoctoral Fellow with the University of California, Los Angeles, in 1985–1986. In 1986–1997, he was with Telstra Research Laboratories (TRL), first as a Research Engineer and, in 1988–1997, as a Project Leader. He also taught and supervised graduate students at Monash University in 1990–2001. During 1997-2008, he was with The University of Melbourne, Victoria, Australia. In 2008 he joined City University of Hong Kong as a Chair Professor of Information Engineering, and a team leader. He has over 300 publications in scientific journals and conference proceedings. He has served on various editorial boards such as Computer Networks, IEEE Communications Magazine, IEEE Journal of Selected Areas in Communications, IEEE/ACM Transactions on Networking and Computer Communications.
\end{IEEEbiographynophoto}

\vfill
\end{document}